# Active terahertz modulator and slow light metamaterial devices with hybrid graphene-superconductor photonic integrated circuits

Samane Kalhor[1], Stephan J. Kindness[2], Robert Wallis[2], Harvey E. Beere[2], Majid Ghanaatshoar[3],

Riccardo Degl'Innocenti[4], Michael J. Kelly[2,5], Stephan Hofmann[5], Charles G. Smith[2],

Hannah J. Joyce[5], David A. Ritchie[2], and Kaveh Delfanazari[1,2,5,*]

[1] *James Watt School of Engineering, University of Glasgow, Glasgow G12 8QQ, UK*

[2] *Cavendish Laboratory, University of Cambridge, Cambridge CB3 0HE, UK*

[3] *Laser and Plasma Research Institute, Shahid Beheshti University, G.C., Evin, 1983969411 Tehran, Iran*

[4] *Department of Engineering, University of Lancaster, Bailrigg, Lancaster, LA1 4YW UK*

[5] *Engineering Department, University of Cambridge, Cambridge CB3 0FA, UK*

* Corresponding Author: kaveh.delfanazari@glasgow.ac.uk

Dated: 14062021

**ABSTRACT:** Metamaterial photonic integrated circuits with arrays of hybrid graphene-superconductor coupled split-ring resonators (SRR) capable of modulating and slowing down terahertz (THz) light are introduced and proposed. The hybrid device optical responses, such as electromagnetic induced transparency (EIT) and group delay, can be modulated in several ways. First, it is modulated electrically by changing the conductivity and carrier concentrations in graphene. Alternatively, the optical response can be modified by acting on the device temperature sensitivity, by switching Nb from a lossy normal phase to a low-loss quantum mechanical phase below the transition temperature ($T_c$) of Nb. Maximum modulation depths of 57.3 % and 97.61 % are achieved for EIT and group delay at the THz transmission window, respectively. A comparison is carried out between the Nb-graphene-Nb coupled SRR-based devices with those of Au-graphene-Au SRRs and a significant enhancement of the THz transmission, group delay, and EIT responses are observed when Nb is in the quantum mechanical phase. Such hybrid devices with their reasonably large and tunable slow light bandwidth pave the way for the realization of active optoelectronic modulators, filters, phase shifters, and slow light devices for applications in chip-scale quantum communication and quantum processing.

■ **INTRODUCTION**

Metallic superconductors are macroscopic quantum systems and gain their electromagnetic properties from pairs of electrons, Cooper pairs[1]. Due to their intrinsic low-loss and plasmonic properties, they are excellent platforms for applications, especially in cryogenic nano-electronics and nano-photonics[2]. Graphene is a thin layer of carbon atoms arranged in a hexagonal network. It is a two-dimensional (2D) material, the thinnest example of a material[3]. The combination of 2D materials and superconductors offers novel electronic and photonic properties that may not be found in either of these materials independently[4,5]. For example, it is possible to measure the superconducting gap in graphene when it is placed in close and clean proximity to a host superconducting material such as niobium (Nb), via the proximity effect[6]. A proximitized graphene then may show exotic electronic or photonic properties, such as topological phases, that are robust against weak external perturbation and are proposed for applications in fault-tolerant quantum computing[7].

The surfaces of hybrid materials can be engineered to achieve electromagnetic response at will. Such engineered materials, metamaterials, develop their exotic electromagnetic properties from the geometries of the engineered unit cells (artificial atoms), through the interactions of artificial atoms and photons[8]. For example, electromagnetically induced transparency (EIT) has been found to have a classical analogy using metamaterials[9]. EIT is a nonlinear quantum effect that evolves due to the destructive interference between excitation states in three-level atomic systems, which results in a narrow transparency window in the medium where light can pass through without any absorption[10]. Classical analogue of the EIT observable in integrated metamaterial devices offers an extreme modification of dispersion properties resulting in the group delay enhancement, slow down, and storage of light[9,11].

An integrated optoelectronic device with the capability of continuous tuning and controlling of the group velocity of light is therefore of interest in the microwave, millimeter wave, terahertz (THz), and optics which make possible for example, the realization of (i) tunable optical phase shifter, (ii) time-delay lines[12] as tools to control the emission of the optoelectronic and telecommunication systems, (iii) continuously tunable fast bandpass filter or bandstop filter[13]. Such devices are also a very useful tool for actively controlling the dispersion, as external cavity mirrors, for ultrashort pulsed QCL[14–16]. Moreover, they can be used for active modulation and polarization control of superconducting IJJ THz quantum emitters[17-29]. The demonstration of the EIT effect in 'static' superconducting THz metamaterials was reported where the authors observe tuning response with intense THz light[30], temperature[9,31,32], and applied DC bias voltage[33]. In this work, we introduce the first class of 'active' and tunable THz modulators and slow light devices based on hybrid graphene-superconductor quantum meta-circuitry.

### ■ DESIGN OF THE HYBRID PHOTONIC INTEGRATED CIRCUITS

THz device offering asymmetric couplings within a pair of static gold split ring resonators (SRRs) was reported by Chen *et al*[34]. Kindness *et al* later demonstrated active control of EIT in such THz metamaterial devices by combining gold-based SRRs with graphene at frequencies above 1.2 THz[35]. In this work, a single layer of graphene is integrated with niobium (Nb) plasmonic SRRs in a superconducting circuit, to realize a voltage and temperature tunable slow light opto-electronic device operating at cryogenic temperatures. The architecture of the hybrid graphene-superconductor photonic integrated circuit is schematically shown in Figure 1(a). The device contains a two-dimensional array of coupled SRRs), oppositely facing each other, in a superconducting circuit.

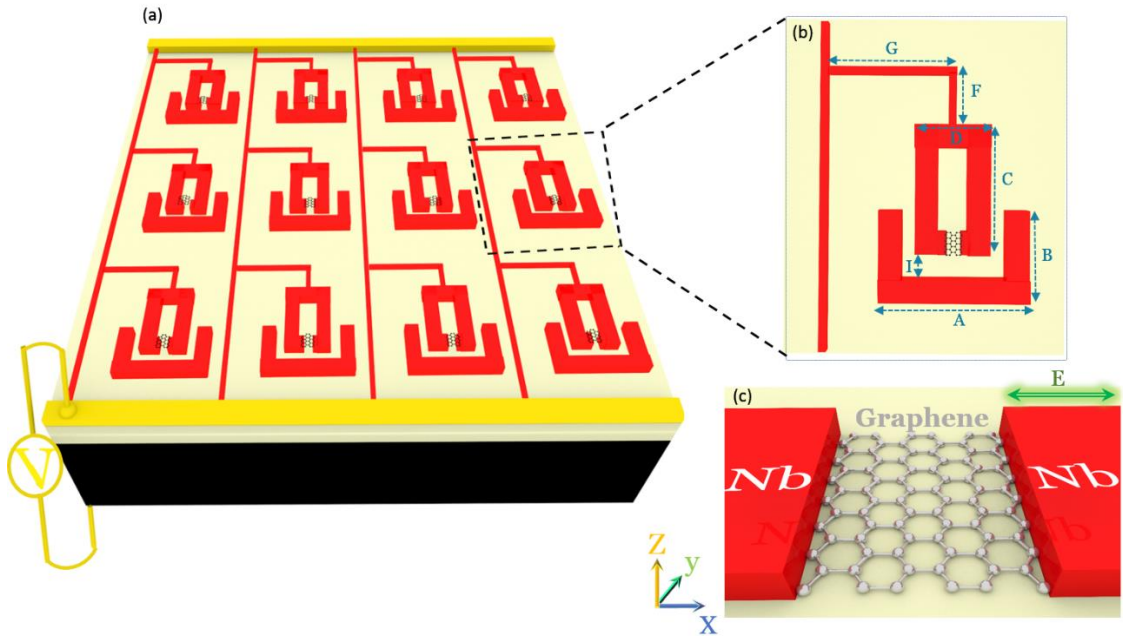

**Figure 1. Voltage-controlled hybrid graphene-superconductor THz photonic integrated circuit.** (a) The hybrid device architecture showing arrays of coupled graphene-Nb SRR integrated with superconducting circuits. Here, Nb, Au, $SiO_2$, Si, and graphene are shown in red, gold, cream, black, and gray, respectively. (b) A meta-atom (unit-cell of the device). (c) The magnified gap structure of a meta-atom, showing graphene is sandwiched between two Nb fingers in the upper resonator in the superconducting circuit. Polarization of the incident THz wave is shown by a green arrow.

A single layer of graphene is integrated with the top resonator, in its small finger (capacitor gap). The superconducting material is considered to be Nb. The THz slow light device with a tunable THz superconductor-graphene-superconductor resonator can be modulated (i) thermally: by switching from a lossy metallic to a low-loss superconducting medium, due to the increase in the Cooper-pairs densities below the transition temperature ($T_c$) of Nb, and (ii) electrically: by changing the conductivity and carrier concentrations in graphene.

Figure 1(b) shows a meta-atom in the device, with dimensions set to be as following: $A$= 52 μm, $B$= 42.4 μm, $C$= 46 μm, $D$= 26.8 μm, $F$= 15.6 μm, $G$= 39.6 μm, $I$= 8 μm, the upper SRR gap length of 4 μm (distance between two Nb fingers), the width of 8 μm, and bias line width of 4 μm. The unit cell periodicity is set to be 114.4 μm × 76 μm. The single unit cell is repeated into arrays to ease light coupling. The thickness of Nb is considered to be 100 nm. Figure 1(c) depicts the magnified gap structure of the top resonator, showing graphene is sandwiched between two Nb fingers of the top SRR. The polarization of the incident THz wave is shown with a green arrow in Figure 1(c). The device is designed based on a 500 μm thick boron p-doped silicon Si substrate with a 300 nm insulating layer of $SiO_2$. The THz transmission of bare $SiO_2$/Si substrate is used as a reference. A Drude model is used for the conductivity of graphene. For describing the variation of graphene conductivity only the DC conductivity $\sigma_0$ is mentioned in the paper. The detailed conductivity of graphene, superconductor Nb, gold, and substrate permittivity are described in Appendix A1.

## ∎ RESULTS AND DISCUSSION

In order to design and model the device architecture, the RF module of COMSOL Multiphysics 5.5 is used. We first consider the THz response of three sets of static superconducting SRRs in the absence of graphene, at $T$= 4.5 K far below the transition temperature $T_c$ of Nb. We consider SRRs with geometries shown in Figure 2(b), here called the bottom single resonator, Figure 2(c), the top single resonator, and Figure 2(d), as coupled resonators that are faced oppositely (rotated by 180 degrees with respect to each other). The THz transmission of these SRRs designs are shown in blue, green, and red lines in Figure 2(a), respectively. In the absence of coupling between two resonators, each SRR independently supports localized surface plasmon (LSP) resonances [13,35]. They show a typical inductive-capacitive (LC) resonance, with center frequencies at 0.53 THz (labeled $f_b$) and 0.58 THz (labeled $f_d$), respectively.

The bottom resonator's longer side is directly excited by the incoming THz $E$-filed. Therefore, the excited net dipole moments in the longer arm result in a resonance with a broader full width at half maximum bandwidth (FWHM), as the radiation damping is SRR structure size-dependent. The total length of the bottom resonator is also slightly larger than the top resonator resulting in resonance at a lower frequency.

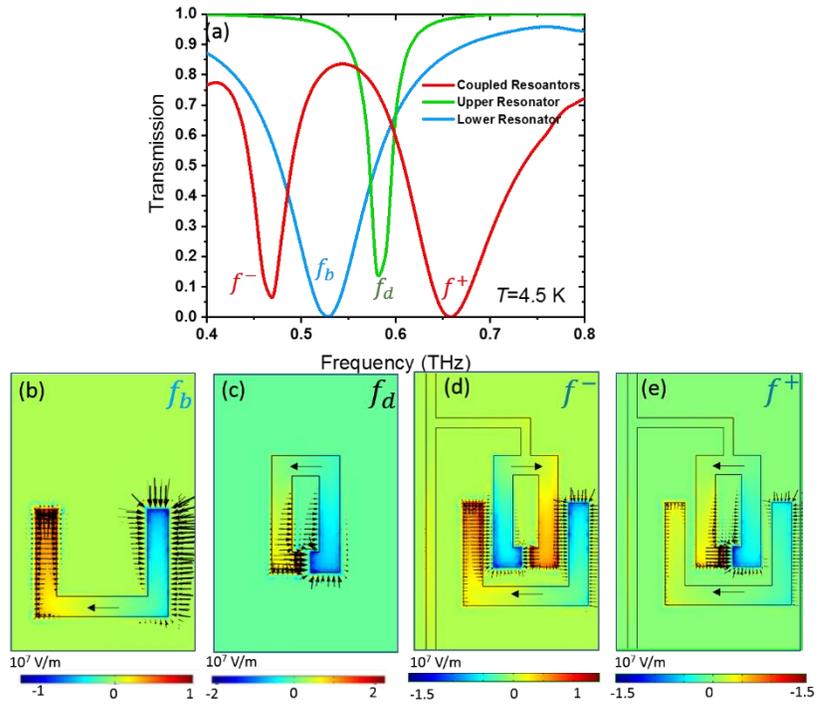

**Figure 2.** (a) Simulated THz transmission of static superconducting Nb SRR arrays in the absence of graphene. Transmission curves when only bottom ring structure (blue), when only top ring structure (green), and when two coupled ring resonators (red) are used at $T= 4.5$ K. (b)-(e) The $z$-component of the electric field $E_z$ and surface current distribution at relevant resonance frequencies, shown in (a).

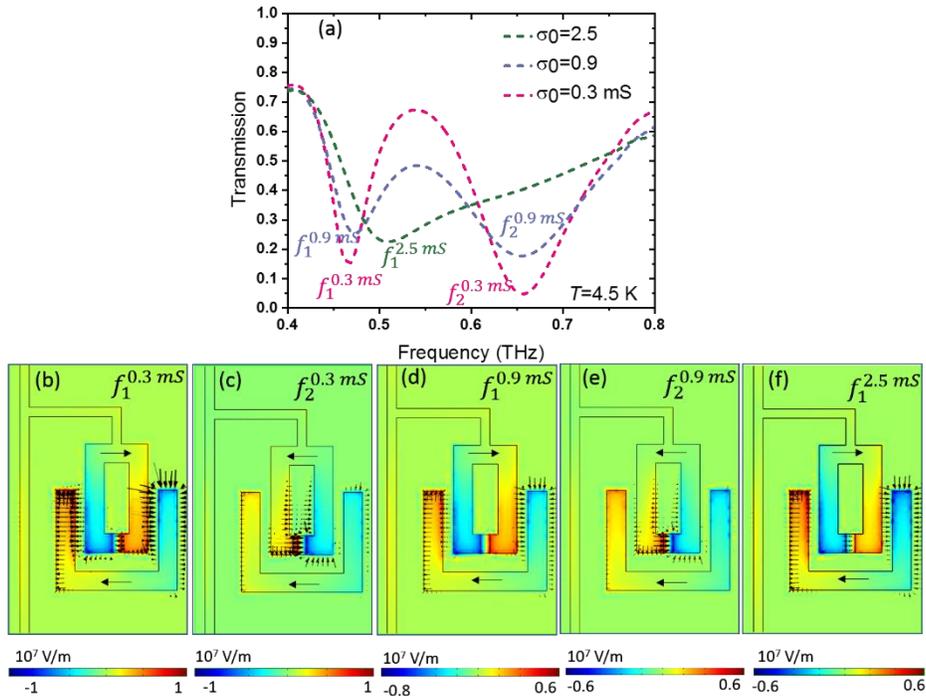

**Figure 3.** (a) Simulated THz transmission of the active/electrically tunable photonic integrated circuit with hybrid graphene-superconducting Nb SRR arrays for three different graphene conductivity, $\sigma_0=0.3$ mS (dashed-red), $\sigma_0=0.9$ mS (dashed-purple), and $\sigma_0=2.5$ mS (dashed-green). (b)-(f) The $z$-component of the electric field $E_z$ and surface current distribution at relevant resonance frequencies shown in (a).

The electric field distribution in the coupled resonators shows the strength of the resonators to incident THz radiation polarized along the *x*-direction, see Figure 2(b)-(e). The smaller SRR (top) acts as a subradiant (or dark) resonator with resonance frequency $f_d$. On the other hand, the larger SRR (bottom) acts as a superradiant (or bright) resonator with resonance frequency $f_b$. However, the destructive interference between two subradiant and superradiant resonances leads to two dips as hybridized modes of "bonding" and "anti-bonding" modes and one peak as asymmetrically coupled-mode[36], as shown with the red curve in Figure 2(a). The electric field and current distribution at the bonding $f^-$ and anti-bonding $f^+$ resonance frequencies are shown in the lower panel of Figure 2. The induced current directions in two coupled SRR are opposite for $f^-$ bonding mode, while they are in the same direction for $f^+$ anti-bonding mode. The suppressed currents due to strong asymmetric coupling and destructive interference between a subradiant mode from the top SRR and a superradiant mode from the bottom SRR lead to induced electromagnetic transparency (EIT) in the coupled SRRs[37,38].

Now, we discuss the results for the active, electrically tunable device, based on the integration of graphene with Nb SRR arrays in a superconducting circuit. The preliminary graphene characterization of the continuous patch was done at room temperature and reported in Appendix Figure A1, providing us with a range of conductivity that we assume for our MM arrays as well. The THz transmission spectra of the device with graphene in the finger of the top SRR, with lowest $\sigma_0$= 0.3 mS (close to Dirac point), $\sigma_0$= 0.9 mS (corresponding to gate voltage $V_{BG}$= 0 V), and highest $\sigma_0$=2.5 mS conductivity values are shown in dashed red, purple and green, in Figure 3(a). By comparing Figure 2(a) and Figure 3(a), one can see that the integration of graphene with dark resonator results in the increase of damping in the resonance of the dark resonator. The results suggest that by the electrostatic gating of graphene one can actively tune the device resonance frequency between a strongly coupled-resonator circuit with superradiant resonance $f_1^{0.3\,mS}$, and a single resonator LSP circuit with sub-radiant resonance $f_1^{2.5\,mS}$. Figure 2 and Figure 3 also show the blueshift of the $f^-$ mode by 41 GHz when the graphene conductivity changes from $\sigma_0$=0 mS, a static resonator circuitry with no graphene in the dark resonator's gap, to $\sigma_0$=2.5 mS where the graphene conductivity is set to highest value ($f_1^{2.5\,mS}$ mode shown in dashed-green). The electric field distribution of this mode shows the damping of the resonance at the dark element (see lower panel of Figure 3).

The device is designed in a way to have electric fields concentrated in the bright and dark resonators, despite the introduction of the superconducting circuit, that is used to connect the dark resonators in the device and to bias the graphene patch (see Appendix A2). The circuit exhibits a resonance at around 0.2 THz so with less impact on the THz transmission response of the coupled resonator arrays. To get further insight into the voltage-controlled active hybrid graphene-superconductor photonic integrated circuit demonstrated in this paper, we compare the detailed results for the case (i) when Nb is in superconducting quantum phase state below $T_c$, and (ii) when Nb is in the resistive normal state above $T_c$. The THz transmission spectra as a function of frequency for different graphene conductivity, between $\sigma_0 = 0$ mS and $\sigma_0 = 2.5$ mS, at $T = 4.5$ K $< T_{c\,Nb}$, and at $T = 10$ K $> T_{c\,Nb}$ are shown in Figure 4(a) and (b), respectively. Figure 4(c) and (d) show the resonance frequencies and transmission dips at resonance frequencies of the device, as a function of graphene conductivity at two different temperatures $T = 4.5$ K, and $T = 10$ K.

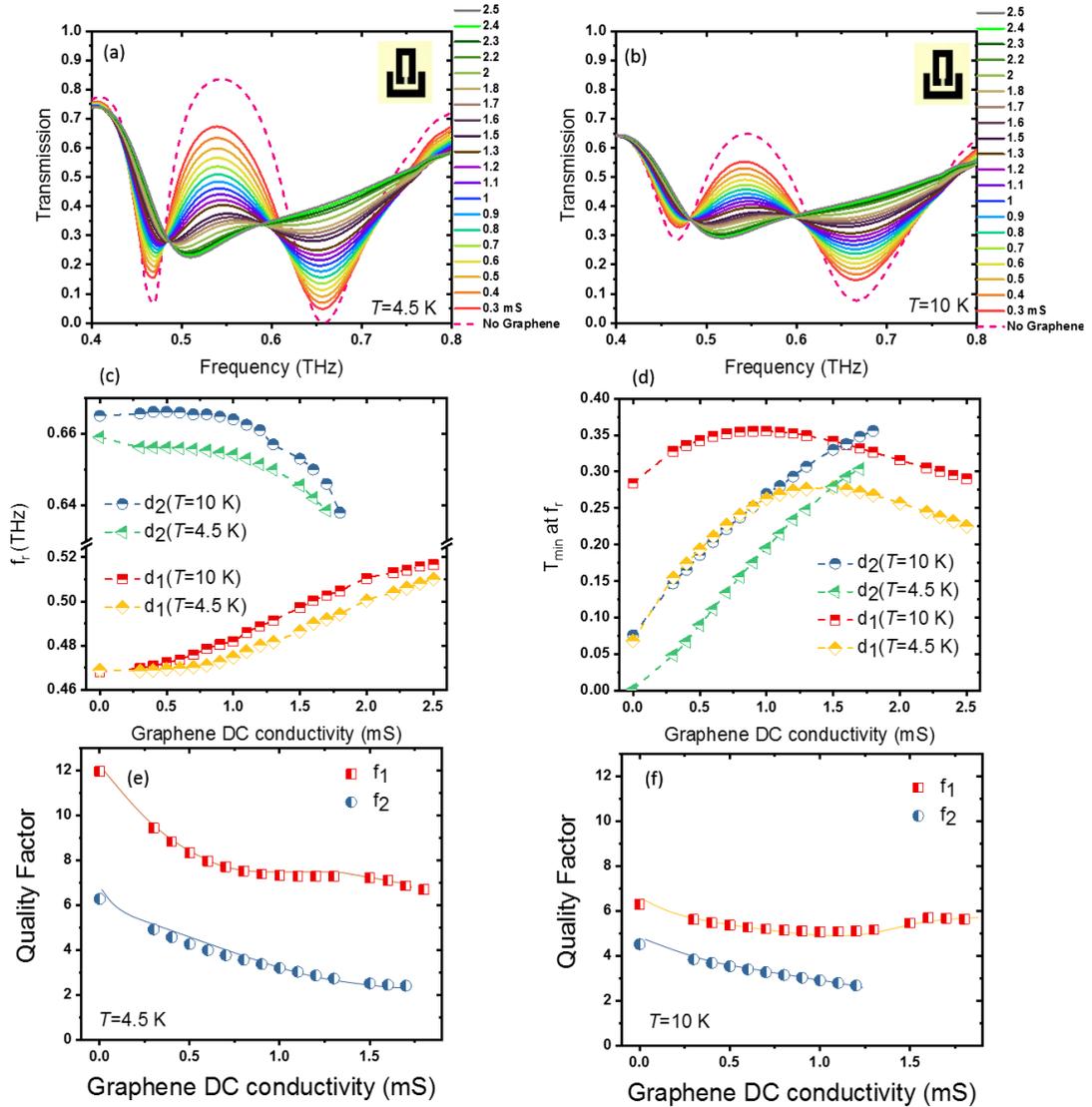

**Figure 4.** THz transmission of the metamaterial photonic integrated circuit with hybrid graphene-Nb SRRs, for different graphene conductivities, (a) at $T = 4.5$ K, far from the Nb superconducting transition temperature $T_c$ where Nb is in superconducting quantum phase, and (b) at $T = 10$ K, where Nb is no longer in superconducting quantum phase. (c) Resonance frequencies and (d) Transmission dips at resonance frequencies of the metamaterial integrated circuit, as a function of graphene conductivity at $T = 4.5$ K, and $T = 10$ K, (e) Quality factor at resonance frequencies at $T = 4.5$ K and (f) Quality factor at resonance frequencies at $T = 10$ K.

A clear enhancement of THz transmission, superradiant and subradiant modes, and frequency tuning can be seen when the device is in superconducting quantum phase state at $T=4.5$ K. Quality factor of first ($f_1$) and second ($f_2$) resonances for the temperatures $T=4.5$ K and $T=10$ K are shown in Figure 4(e) and (f), respectively. With increasing the graphene DC conductivity, the strength of both resonances decreases up to the $\sigma_0=1.8$ mS where the second resonance completely damps and only one single resonance remains.

We also compare the effect of the superconducting quantum phase on the THz response of such active devices, with those of devices in the normal phase based on different materials. Figure 5 summarizes the THz transmission of the hybrid photonic integrated circuit, when the bright/dark SRR resonators are set to be as: (a) Nb (at $T=4.5$ K)/Au, (b) Nb (at $T=10$ K)/Au, (c) Au/Nb (at $T=4.5$ K), (d) Au/Nb (at $T=10$ K), and (e) Au/Au, respectively. One can clearly find the role of superconductivity. Looking at Figure 5(a)-(e), we find that the strongest response, in both resonance quality and EIT peak, is observed when both coupled resonators are made from Nb and when the Nb is set at $T=4.5$ K $< T_{cNb}$. This is because at this temperature the THz surface resistance of Nb is significantly small, a value around $R_s=100$ m$\Omega$ between $f=100$ GHz and 600 GHz, at $T=5$ K[12]. Replacing Nb in the bright resonator with Au will reduce the response (increase the FWHM) but this reduction is significant when Nb in the dark resonator is replaced by Au (see Figure 5(f)). The device transmission minimum at resonance frequencies will significantly weaken when the SRR structure is built from Au as shown in Figure 5(g)-(h), which retains large surface resistance at all temperatures.

We use an equivalent lumped element circuit model to further investigate the resonance response and frequency tuning of the coupled resonators as a function of graphene conductivity. Figure 6(a) shows the $RLC$ electrical circuit equivalent of the SRR. In the absence of coupling, each single SRR can be considered as an antenna with a frequency-dependent impedance. The electric circuit in Figure 6(a), with the lower loop consisting of a resistance $R_1$, an inductance $L_1$, and a capacitance $C_1$ represents the bright resonator. The upper loop represents the dark resonator with $R_2$, $L_2$, and $C_2$ circuit elements. In the presence of coupling, due to the electric field of charges in close proximity between two bright and dark SRRs, a parallel coupling capacitor $C_c$ can be considered to connect the two circuit loops. In the circuit, $R_1$ and $R_2$ illustrate the losses in each SRR, $L_1$ and $L_2$ represent the stored magnetic energy due to induced current in each SRR, $C_1$ and $C_2$ illustrates the energy stored in the finger of each bright and dark SRR due to accumulated charges. $R_G$ accounts for the extra induced (resistive) losses as a result of the integration of graphene with Nb in the finger of dark SRR. $V_1$ and $V_2$ in the circuit represent the electromotive force on the electrons due to the incident THz radiation in the bright and dark SRR, respectively. We assume that the value of $V_2$ is 0.65 $V_1$ as the bright SRR provides a larger coupling with the incident THz transmission. The inductance of each resonator $L_i$ is defined as the ratio of magnetic field flow divided by current $J$ density, as $L_i = \frac{\int H_z dxdy}{\int J dxdy}$. The $z$-component of magnetic field $H_z$ and induced current density $J$ for each SRR are obtained from the COMSOL simulation. After determining the inductance of each SRR, the capacitance is calculated from the equation $\omega_r = \frac{1}{\sqrt{L_i C}}$ [27,39].

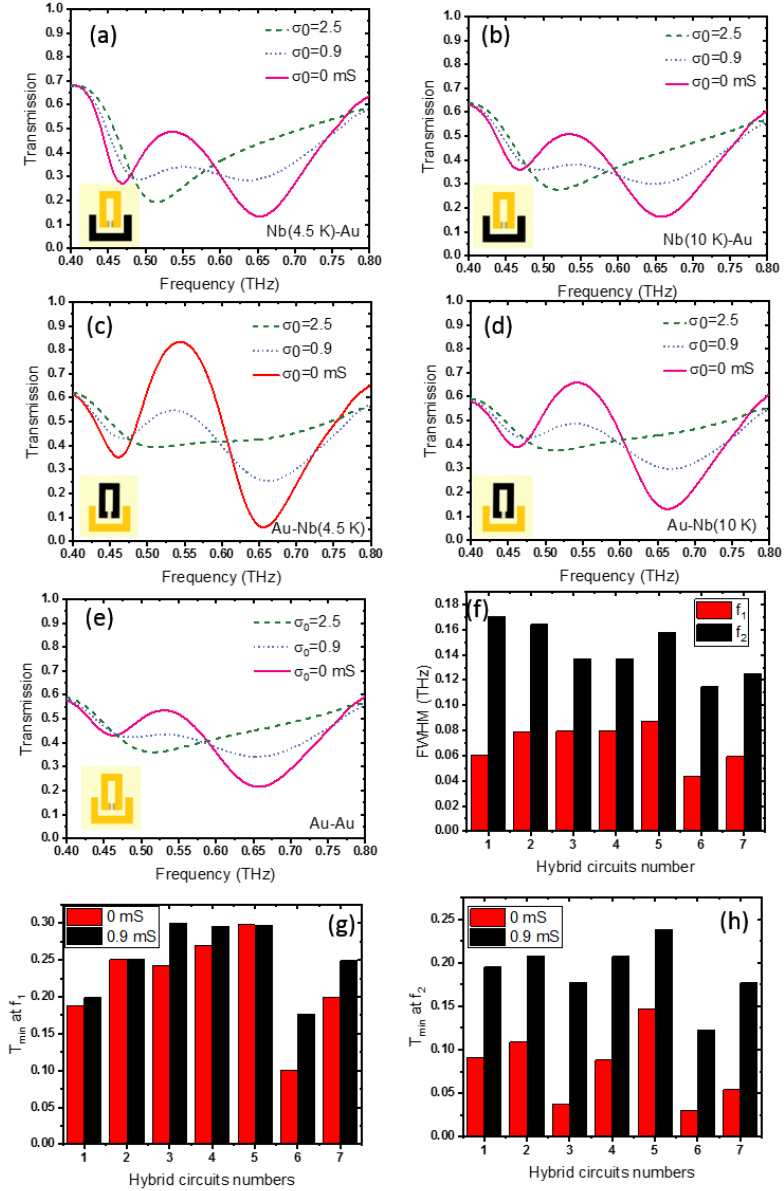

**Figure 5.** THz transmission of the hybrid photonic integrated circuit when the bright/dark resonators are set to be (a) Nb (at $T=4.5K$)/Au and (b) Nb (at $T=10K$)/Au, at three different graphene conductivity. Same info is given in (c) and (d) but with Nb and Au replaced in the SRR unit-cells. Insets show one unit-cell where Nb SRR is shown black, Au SRR is shown yellow. (e) THz transmission of the photonic integrated circuit when the bright/dark resonators are both set to be Au. (f) FWHM at $\sigma_0=0$ mS, (g) transmission minimum at first resonance, and (h) transmission minimum at second resonance of the hybrid circuit bright/dark resonators. Here, each device is labeled as 1- Nb (at $T=4.5K$)/Au, 2- Nb (at $T=10K$)/Au, 3- Au/Nb (at $T=4.5K$), 4-Au/Nb (at $T=10K$), 5-Au/Au, 6- Nb (at $T=4.5K$)/ Nb (at $T=4.5K$), 7- Nb (at $T=10K$)/ Nb (at $T=10K$) at $\sigma_0=0$ mS and at $\sigma_0=0.9$ mS.

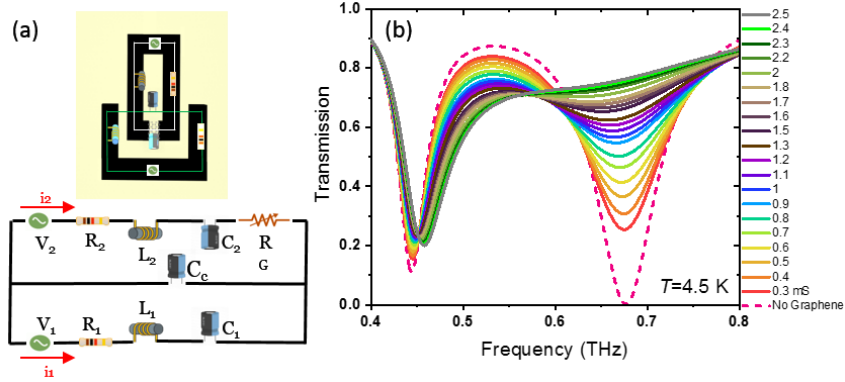

**Figure 6.** (a) The equivalent RLC circuit model of the hybrid graphene-Nb coupled resonators. (b) THz transmission and frequency tuning of the hybrid SRR for different graphene conductivity, between $\sigma_0=0$ and $\sigma_0=2.5$ mS, obtained by using equivalent RLC model.

A free fitting parameter as $C_c = 20\ fF$ is used to obtain $C_1$ and $C_2$ from the equation $C = \frac{C_i + C_c}{C_i C_c}$. The resistance $R_i$ of each SRR is calculated from the equation $Q_i = \frac{1}{R_i}\sqrt{\frac{L_i}{C_i}}$ [40], where the quality factor of each ring is obtained from $Q_i = \frac{f_r}{\Delta f}$. Here, $f_r$ and $\Delta f$ are resonance frequency and FWHM of the resonance, respectively [27]. The obtained circuit element values for each loop are shown in table 1. The dampening resistor $R_G$ as a function of graphene conductivity $\sigma_0$ is determined from the equation $Q = \frac{1}{R_{no} + R_G}\sqrt{\frac{L_2}{C_2}}$.

First, the quality factor of the resonance of a single dark SRR is determined by performing a COMSOL simulation. Then the resistance of the dark SRR without graphene ($R_w$) is calculated. $R_G$ changes from 2.5 Ω for $\sigma_0= 0.3$ mS to 25 Ω for $\sigma_0= 2.5$ mS. The current induced in the bright and dark SRR (named as $i_1$ and $i_2$ respectively) will flow through the bottom and top loops. The voltages are related to currents through Kirchhoff's law (see Appendix equation A1). Finally, transmission $T$ can be obtained from $T = 1 - \frac{1}{2}(Re\ (V_1 i_1^* + V_2 i_2^*))$ where the complex conjugate of the current induced in the bright and dark resonators is described by $i_1^*$ and $i_2^*$, respectively [35,41].

The modeled THz transmission of the hybrid graphene-Nb SRR, as a function of frequency for different graphene DC conductivity (and also for when there is no graphene in the finger of the dark SRR) at $T= 4.5$ K, is shown in Figure 6(b). The electric field $E_{gap}$ across the capacitive gap of the Nb SRR and the graphene conductivity will determine the power dissipation and therefore the amount of resonance dampening. As $E_{gap}$ increases, $R_G$ increases and the second resonance quality decreases. One can see that the results of the THz transmission obtained from the RLC equivalent circuit model based on destructive interference between $i_1$ and $i_2$ are in reasonably good agreement with COMSOL simulation (See Figure 4(a)). Next, we focus on the EIT peak which is observed between two bonding $f^-$ and anti-bonding $f^+$ resonances (see Figure 2). The EIT offers an extreme modification of the dispersion, resulting in the slow light effect. Such slow light effect in hybrid metamaterial integrated circuits can be observed as a result of the group velocity alteration in the device.

**TABLE 1:** The circuit parameters for lower and upper loops.

| $R_1$ (Ω) | $L_1$ (pH) | $C_1$ (fF) | $R_2$ (Ω) | $L_2$ (pH) | $C_2$ (fF) |
|---|---|---|---|---|---|
| 4.2712 | 0.2156 | 7.2453 | 6.9322 | 0.1276 | 5.9021 |

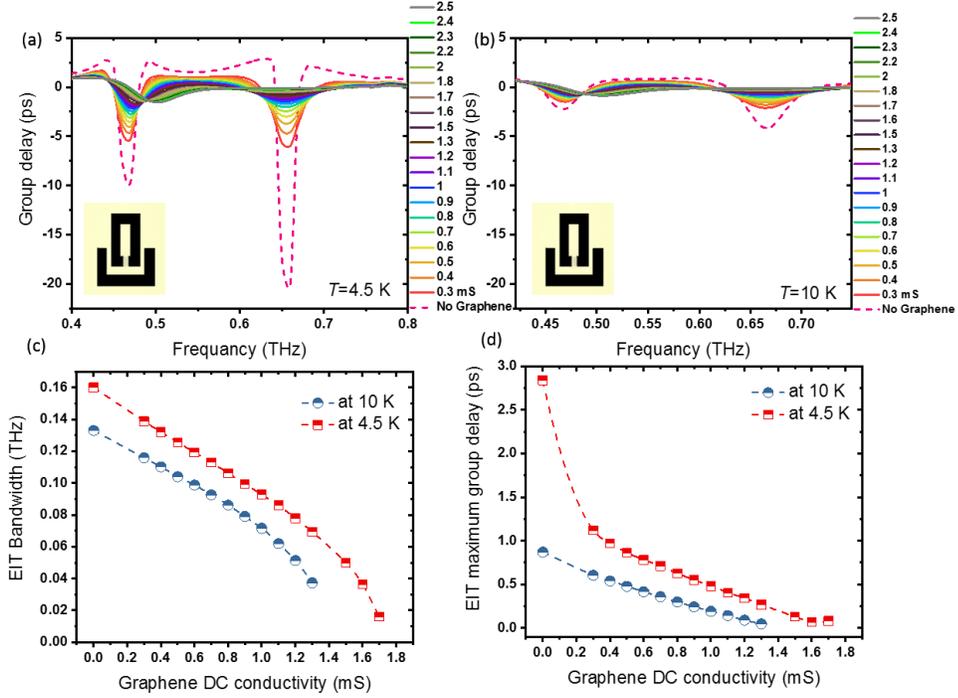

**Figure 7.** Group delay through the hybrid photonic integrated circuit for different graphene conductivity, at (a) $T$= 4.5 K, far from the Nb superconducting transition temperature $T_c$, and (b) at $T$= 10 K, where Nb is no longer in superconducting quantum phase. (c) the bandwidth of transparency window, and (d) EIT maximum group delay through the metamaterial photonic integrated circuit, with hybrid graphene-Nb SRRs, as a function of graphene conductivity at $T$= 4.5 K < $T_c$, and at $T$= 10 K > $T_c$.

Here, we demonstrate how to actively control slow light in hybrid graphene-superconductor resonators integrated with superconducting circuits. The group delay $t_g$, the time delay of a THz wave packet traveling through the device in comparison to air, is determined by $t_g = \frac{d\phi}{d\omega}$, where $\phi$ is the phase of the THz transmission and $\omega$ is the angular frequency. Group delay is calculated relative to the bare substrate reference. Figure 7 shows the group delay through the device as a function of graphene conductivity for temperatures (a) $T$= 4.5 K < $T_c$ and (b) $T$= 10 K > $T_c$. Maximum negative group delay dips are observed at around 0.46 THz and 0.66 THz at both temperatures, however, the group delay value and shape observed at $T$= 4.5 K < $T_c$, where Nb is in the superconducting quantum phase, is much larger and sharper. The positive group delay, between these frequencies, indicates the slow light effect through the hybrid device. It can be seen that the delay time is quite loss sensitive and is much larger when Nb is in the superconducting quantum phase at $T$=4.5 K. A maximum positive group delay of 2.86 ps at 0.627 THz is observed when there is no graphene present in the Nb gap in the dark resonators (with Nb in its smallest THz surface resistance value). The maximum positive slow light and the group delay values decrease as the conductivity of graphene increases (damping increases) for both temperatures. Figure 7(c) and (d) summarize the bandwidth of transparency window and EIT maximum group delay through the metamaterial photonic integrated circuit, as a function of graphene conductivity at $T$= 4.5 K, and at $T$= 10 K. Our results indicate that the proposed hybrid photonic integrated circuit based on graphene-Nb SRR can be used as an efficient and tunable THz slow light modulator device.

Applying a voltage on the device pads can significantly change the amplitude of the EIT peak, rather than the resonance frequencies, as is clear from Figure 7(a)-(b). To quantitatively investigate this variation, we introduced the modulation depth (MD) of EIT amplitude as a function of graphene conductivity as $MD_{amplitude} = \left|\frac{T_{(0\,mS)} - T}{T_{(0\,mS)}}\right|$, where $T_{(0\,mS)}$ is transmission at $\sigma_0$=0 mS, and $T$ is transmission at the desired graphene conductivity. Both $T_{(0\,mS)}$ and $T$ are obtained at the same frequency of $f$=0.541 THz. The MD of EIT amplitude (shown in Figure 8) increases with graphene conductivity. It has a maximum value of 57.3 % at $\sigma_0$=1.6 mS for $T$=4.5 K. With increasing the graphene conductivity, the EIT transmission has an obvious decline that results in MD growing. When $\sigma_0$ comes to larger than 1.6mS at $T$=4.5 K, the EIT peak shows no more positive group delay. Positive group delay corresponds to the slowness of slow light devices. In addition, active modulation of slow light can be realized by graphene conductivity. The modulation depth of group delay is determined by $MD_{group\,delay} = \left|\frac{t_{g_{(0\,mS)}} - t_g}{t_{g_{(0\,mS)}}}\right|$. MD of slow light shown in red in Figure 8 shows an increase with graphene conductivity. MD of group delay is as high as 97.61 % at $\sigma_0$=1.6 mS for $T$=4.5 K.

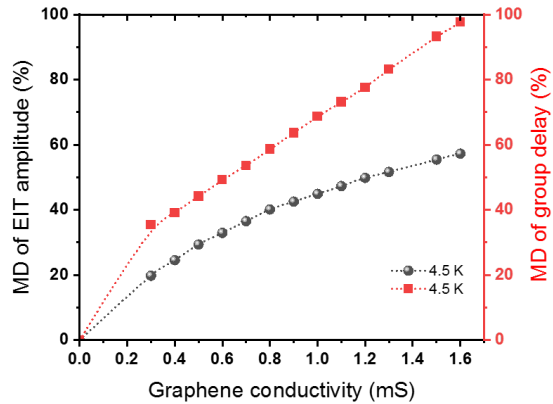

**Figure 8.** Modulation depth (MD) of the EIT and group delay as function of graphene conductivity at $f$=0.541 THz, and $T$=4.5 K.

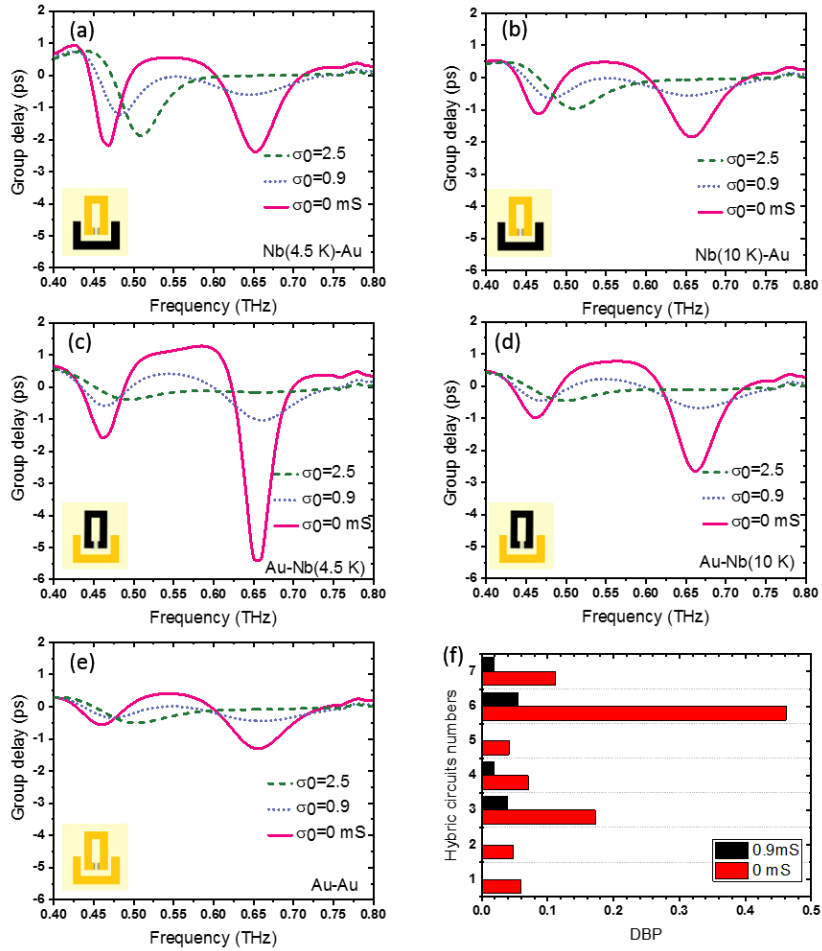

**Figure 9.** Group delay through the hybrid photonic integrated circuit for different graphene conductivity when the bright/dark resonators are set to be (a) Nb (at $T= 4.5$K)/Au and (b) Nb (at 10K)/Au. Same info is given in (c) and (d) but with Nb and Au replaced in the SRR unit-cells. Insets show one unit-cell where Nb SRR is shown black, Au SRR is shown yellow. (e) Group delay through the hybrid photonic integrated circuit for different graphene conductivity when the bright/dark resonators are both set as Au. (f) delay-bandwidth product of hybrid device bright/dark resonators for device # 1- Nb (at $T= 4.5$K)/Au, 2- Nb (at $T= 10$K)/Au, 3-Au/Nb (at $T= 4.5$K), 4-Au/Nb (at $T= 10$K), 5-Au/Au, 6- Nb (at $T= 4.5$K)/ Nb (at $T= 4.5$K), 7- Nb (at $T= 10$K)/ Nb (at $T= 10$K), at $\sigma_0=0$ mS and $\sigma_0=0.9$ mS.

Figure 9, with coupled SRR of different materials, provides further insight into the effect of the superconducting quantum phase on the group delay and slow light effect in such active devices. Here, the bright/dark SRRs are set as: (a) Nb (at $T= 4.5$ K)/Au, (b) Nb (at $T= 10$ K)/Au, (c) Au/Nb (at $T= 4.5$ K), (d) Au/Nb (at $T= 10$ K), and (e) Au/Au. One can clearly find the role of the superconducting quantum phase state. The strongest EIT response and slow light effect are observed when both coupled resonators are based on Nb, and when the Nb is set at $T= 4.5$ K $< Tc_{Nb}$. This is because at this temperature the THz surface resistance of Nb is much smaller. Replacing Nb in the bright resonator with Au will reduce the group delay but this reduction is considerable when Nb in the dark resonator is replaced by Au. The device response will significantly weaken when the SRR structure is built from Au, which retains large surface resistance at $T= 4.5$ K.

The maximum group delay, EIT window bandwidth and maximum delay-bandwidth product (DBP) for different hybrid circuits, including device number 1- Nb (at $T$= 4.5K)/Au, 2- Nb (at $T$= 10K)/Au, 3-Au/Nb (at $T$= 4.5K), 4- Au/Nb (at $T$= 10K), 5- Au/Au, 6- Nb (at $T$= 4.5K)/ Nb (at $T$= 4.5K), 7- Nb (at $T$= 10K)/ Nb (at $T$= 10K), at $\sigma_0$=0 mS and $\sigma_0$=0.9 mS, are calculated and presented in tables A1 of Appendix. DBP for the different hybrid circuits is shown in Figure 9(f). device # 6 (Nb/Nb) at the lowest temperature of $T$=4.5 K have the largest DBP which originated from its low resistance, likewise, device # 5 (Au/Au) has the lowest DBP. Moreover, no EIT characteristic (0 DBP) is observed in the circuit when Nb is replaced by Au in the dark resonator.

■ **CONCLUSIONS**

We proposed a voltage and temperature controlled photonic integrated circuitry by the integration of graphene with an array of Nb subwavelength split ring resonators in a superconducting circuit and reported the first demonstration of a hybrid graphene-superconductor THz metamaterial slow light device. We showed that an equivalent circuit model is a useful tool for active hybrid device optimization. Furthermore, we demonstrated electromagnetic induced transparency, sub-radiant and super-radiant resonances in such novel class of hybrid photonic integrated circuits. The quantitative analysis shows that the modulation depth of EIT transmission amplitude and slow light group delay can be realized which are controllable by voltage and temperature. Our devices with their large and tunable slow light bandwidth pave the way for the realization of active optoelectronic modulators for applications in future quantum communication and computation systems.

■ **ASSOCIATED CONTENT**

# Appendix

### A1) Simulation methods

SiO$_2$ and Si are considered as dielectrics with $\varepsilon$=3.9 and 11.56, respectively[42]. The complex frequency and temperature-dependent conductivity of Nb was extracted from the experimental data[43]. A simple Drude model is used to model the complex conductivity of graphene $\sigma_G(\omega) = \frac{\sigma_0}{1+i\omega\tau}$, where $\tau$= 15 $f$s is the scattering time, and $\omega$ is the angular frequency [35]. The DC conductivity of graphene $\sigma_0$ is set to be between 0.3 mS and 2.5 mS in the simulation. Graphene is simulated as a 2D layer with surface current boundary condition. The surface current of graphene J = $\sigma_G$E is defined as the product of its conductivity ($\sigma_G$) in Siemens unit and the tangential electric field (E). The conductivity of Au ($\sigma_{Au}$) is described by the Drude model expression as $\sigma_{Au} = \varepsilon_0 \frac{\omega_p^2}{\gamma + i\omega}$ where plasma frequency $\omega_p$ is $2\pi \times 2175$ THz and collision frequency $\gamma$ is $2\pi \times 6.5$ THz[44]. Here, $\varepsilon_0$ is the vacuum electric constant.

## A2) Graphene DC conductivity

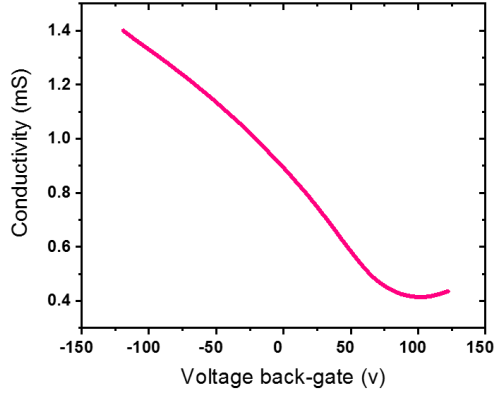

**Figure A1.** The measured conductivity of graphene as a function of back-gate voltage. At zero back-gate voltage ($V_{BG}= 0$) the graphene conductivity is around 0.9 mS, and these values are used in COMSOL simulation [31].

## A3) bias line

The bias line exhibit the resonance at $f$=0.31 THz for $T$=4.5 K. The bias line resonance shows no variation with graphene DC conductivities. The electric field distribution at bias line resonance frequency in Figure A2(b) accumulates at the bias line. On the other hand, the electric field distribution at the peak mode in Figure A2(c) and also two dark and bright resonances (as shown in Figures 2 and 3) show no enhancement at the bias line. Therefore, the bias line is designed to have a low influence on the EIT window between bright and dark resonances.

## A4) equivalent RLC model

The current induced in the bright and dark SRR (named as $i_1$ and $i_2$ respectively) flow through the bottom and top loops can be obtained from the voltage in the bright SRR $V_1$ and the dark SRR $V_2$ based on Kirchhoff's law:

$$\begin{bmatrix} i_1 \\ i_2 \end{bmatrix} = \begin{bmatrix} j\omega L_1 + R_1 + \frac{C_1+C_c}{j\omega C_1 C_c} & \frac{1}{j\omega C_c} \\ \frac{1}{j\omega C_c} & j\omega L_2 + R_2 + R_G + \frac{C_2+C_c}{j\omega C_2 C_c} \end{bmatrix}^{-1} \begin{bmatrix} V_1 \\ V_2 \end{bmatrix} \qquad (A1)$$

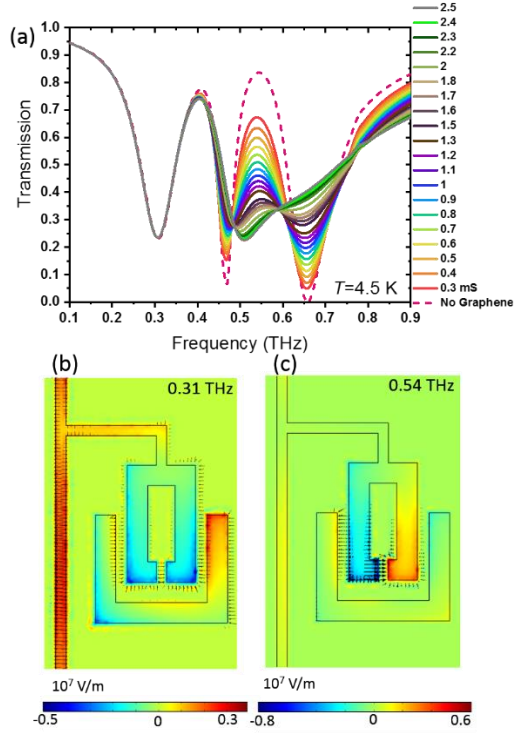

**Figure A2.** (a) Simulation transmission data for *T*=4.5 K at different graphene DC conductivities. Z-component of electric field distribution at $\sigma_0$=0 mS for (b) bias line resonance frequency *f*=0.31 THz, and (c) the peak frequency *f*=0.54 THz

## A5) Group delay characteristics of coupled SRR with different materials

**TABLE A1.** Maximum group delay, EIT bandwidth, and maximum group delay-EIT bandwidth product (DBP) of different hybrid circuits at $\sigma_0$=0 mS.

| @ 0 mS | Device number | Max group delay (ps) | EIT bandwidth (THz) | Max delay-EIT bandwidth product (DBP) |
|---|---|---|---|---|
| **Nb (4.5 K)/Au** | 1 | 0.548 | 0.107 | 0.05863 |
| **Nb (10 K)/Au** | 2 | 0.444 | 0.106 | 0.04706 |
| **Au/Nb (4.5 K)** | 3 | 1.287 | 0.134 | 0.17245 |
| **Au/Nb (10 K)** | 4 | 0.565 | 0.125 | 0.07062 |
| **Au/Au** | 5 | 0.396 | 0.105 | 0.04158 |
| **Nb (4.5 K)/Nb (4.5 K)** | 6 | 2.87 | 0.161 | 0.46207 |
| **Nb (10 K)/Nb (10 K)** | 7 | 0.874 | 0.132 | 0.1118 |

**TABLE A2.** Maximum group delay, EIT bandwidth, and maximum group delay-EIT bandwidth product (DBP) of different hybrid circuits at $\sigma_0$=0.9 mS.

| @ 0.9 mS | Device number | Max group delay (ps) | EIT bandwidth (THz) | Max delay-EIT bandwidth product (DBP) |
|---|---|---|---|---|
| Nb (4.5 K)/Au * | 1 | ------ | ------ | -------- |
| Nb (10 K)/Au * | 2 | ------ | ------ | -------- |
| Au/Nb (4.5 K) | 3 | 0.410 | 0.094 | 0.03854 |
| Au/Nb (10 K) | 4 | 0.210 | 0.086 | 0.01806 |
| Au/Au | 5 | 0.114 | 0.016 | 0.00182 |
| Nb (4.5 K)/Nb (4.5 K) | 6 | 0.555 | 0.099 | 0.05494 |
| Nb (10 K)/Nb (10 K) | 7 | 0.242 | 0.079 | 0.01911 |

*These hybrid circuits show no positive group delay and correspondingly no EIT window at $\sigma_0$=0.9 mS

■ **AUTHOR INFORMATION**

**ORCID:**

Kaveh Delfanazari: 0000-0002-1386-3855

■ **ACKNOWLEDGMENTS**

K.D., M.J.K., H.J.J., D.A.R., C.G.S. acknowledge funding from EPSRC (MQIC). S.H. acknowledges funding from EPSRC (EP/P005152/1). D.A.R., H.B., R.D. acknowledge funding from EPSRC (HyperTHz).

■ **REFERENCES**